# A Showcase of Using the Partial-Structure R1 to Assemble Small-Molecule Crystal Structures

Authors

**Xiaodong Zhang[a]***

[a]Chemistry Department, Tulane University, 6400 Freret Street, New Orleans, Louisiana, 70118, United States

Correspondence email: xzhang2@tulane.edu

**Synopsis** How the partial-structure R1 (pR1) can be used to assemble small-molecule crystal structures is showcased by one concrete example.

**Abstract** Using a few concrete examples this paper has demonstrated a few observations on using the partial-structure R1 (pR1) to assemble small-molecule crystal structures. (1) Assembling can start with orienting and/or placing large fragments or small fragments containing heavy atom(s). (2) It is handy to attach a fragment to a partial model as only the orientation needs to be optimized. (3) In a structure containing heavy atoms it is impossible to directly calculate the orientations of a light-atom-only fragment. Instead, it is necessary to use residual reflection intensities in which the contribution of the partial model containing all heavy atoms is deducted away. These observations indicate that pR1 is a useful tool for solving some small-molecule crystal structures.

**Keywords: partial-structure R1; assembling crystal structure; residual reflection intensities; structure solution**

## 1. Introduction

In modern X-ray crystallography, solving a crystal structure typically relies on phasing (Giacovazzo, 2014), and modern driving force behind phasing is a dual space recycling process, involving reciprocal space phase refinement and/or real space electron density modifications (as pioneered in 1990s: Weeks *et al.*, 1993; Miller *et al.*, 1993; DeTitta *et al.*, 1994). Recently, some methods that do not involve phasing have been explored. One method is based on a diagonal least-squares technique (Burla *et al.,* 2018), and another method relies on optimization of a single-atom R1 (sR1) function (Zhang & Donahue, 2024). Such exploration is mainly for theoretical interest but nevertheless enriches the toolbox for solving crystal structures. The sR1 method can solve small-molecule crystal structures. It was also demonstrated capable of solving a small protein structure after accounting for a hidden solvent diffraction effect (Zhang, 2026).

At the same time when the single-atom R1 (sR1) concept was introduced (Zhang & Donahue, 2024), a more general concept, the partial-structure R1 (pR1) was also introduced. Consider a crystal structure that needs to be solved. Suppose a partial model of this structure has already been determined and one is considering adding a fragment into the model. There are other undetermined atoms besides this fragment. The pR1 function is defined via modifying the traditional R1 by removing its dependence on the location of these other atoms. The resulting approximate R1 (aR1), namely, the pR1 function only depends on the orientation and location of the defining fragment. This is because the coordinates of the atoms in the known partial model are the known parameters. So, in principle, by globally optimizing the pR1 function, the correct orientation and location of the fragment can be determined. Unfortunately, this is a 6-d search, which requires enormous computing power. In practice, it is necessary to divide this 6-d search into two 3-d searches, one for the orientation and the other for the location. In favorable occasions, such division is feasible: for examples, when the fragment is large, or when it contains heavy atom(s). In such cases, the correct orientations can be detected by the local minima of the pR1 function in orientation space if a free-standing form of the fragment is used. Here, free-standing means that the whole model consists of this single fragment only. When a fragment is free-standing, its placement within the cell is immaterial, and so, the pR1 function only depends on the orientation of the fragment. The detected orientations are ordered by the pR1 value from small to large, and are labeled as 0, 1, 2…. Once the orientation of a fragment has been decided, its placement (namely, the placement of the local origin of the fragment) within the cell can now be optimized by minimizing the pR1 function. There is one special case in which the placement of a fragment is known at start, that is, when the fragment is attached to the known partial model. In this case, of course, only the orientation needs to be optimized by a 3-d search. There is one more problem with pR1 calculations. Consider a structure containing heavy atoms and one tries to add a light-atom-only fragment into the model. One tries to use a free-standing form of this fragment to detect its possible orientations. Unfortunately, this trick simply fails. This is because the weak signal of the light-atom-only fragment is overwhelmed by the strong diffraction of the heavy atoms. To find the orientations of a light-atom-only fragment, it is necessary to modify the hkl data by removing the heavy atom contributions to the diffraction intensities. In the following, all these practical considerations for a pR1 calculation are showcased by one concrete example. To make the presentation concise, two additional examples are instead given in the supporting information.

## 2. The example selected for showcase

The dataset code for this example is JPD1252. Its space group is C2/c. To simplify the calculations, the C centered cell has been converted to a primitive cell. In one primitive cell there are four $SiPh_2{}^tBuMoO_4$ fragments and four $N^nPr_4$ fragments. So, this example is well suited to be solved by the pR1 method. An idealized model of the $N^nPr_4$ fragment can readily be set up (see details in the supporting information). However, the structure of the $SiPh_2{}^tBuMoO_4$ fragment is unknown. In the next section, a model of this fragment will be built *in situ* by the pR1 calculations.

### 3. Use pR1 to build a model of the $SiPh_2{}^tBuMoO_4$ fragment *in situ*

To start, consider an idealized model of $MoO_4$ fragment, which has tetrahedral shape with Mo at the center and Mo-O bond length 1.72 Å. Use Mo as the local origin of this fragment, and place it at (0.3, 0.3, 0.3) in the unit cell. The correct orientation of this fragment is then determined by a global minimization of its pR1 function. The resulting model is shown as step 1 in figure 1.

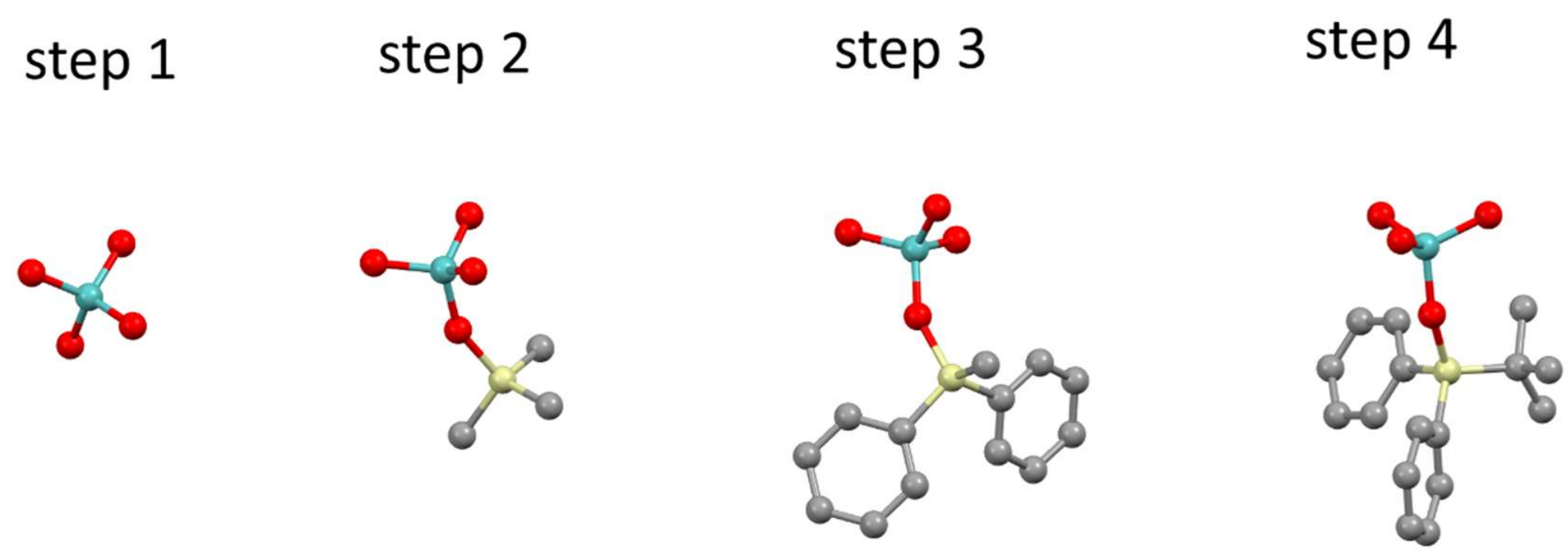

**Figure 1** Steps of building a model of the $SiPh_2{}^tBuMoO_4$ fragment *in situ*

Next, set up an idealized model of $SiC_3O$. It is of tetrahedral shape, with Si at the center and the other atoms at the four corners. The bond length from Si to the other atoms is 1.8 Å. Set the local origin on the O atom. Then delete this O atom from this model. The result is a model of a $SiC_3$ fragment. This fragment needs to be attached to the correctly oriented $MoO_4$ fragment, with its local origin sits on one of the O atoms and its orientation being optimized by minimizing its pR1 function. As there are four O atoms, there are four possibilities. Try all of them and pick the one with the smallest pR1 value. The result is a partial model shown as step 2 in figure 1.

There are three C atoms in the resulting partial model. Two C atoms need to be completed into two benzene rings. Set up an idealized model of a benzene ring with C-C bond length 1.39 Å. Put local origin at one C atom and then delete this atom. The result is a model of a $C_5$ fragment. A $C_5$ fragment can be attached to the partial model with its local origin sitting on one of the three C atoms in the partial model, and its orientation being optimized by minimizing its pR1 function. Two $C_5$ fragments need to be added. There are three possibilities of doing this. Try all of them and pick the one with the lowest pR1 value. The result is a partial model showing as step 3 in figure 1.

The remaining C atom that bonds to the Si atom in the partial model needs to be completed into a ${}^t$Bu group. Set up an idealized model of $C_5$ with a tetrahedral shape and one C at the center and four C atoms at the corners, of C-C bond length 1.55 Å. Set the local origin at the center. Then delete the C at the center and delete one C at one corner. The result is a model of a $C_3$ fragment. Attach one such fragment to the partial model, with its local origin sits on the C atom that bonds to the Si atom. The orientation of this $C_3$ fragment is optimized. The result is a completed model of the $SiPh_2{}^tBuMoO_4$ fragment which is shown as step 4 in figure 1. In the following calculations the local origin of this fragment is on its Mo atom.

#### 4. Detect the correct orientations of the $SiPh_2{}^tBuMoO_4$ fragment

Using a free-standing form of this fragment the possible orientations are detected by the local minima of its pR1 function in the orientation space. These orientations are labeled as 0, 1, 2… in the order from smaller pR1 value to larger pR1 values.

#### 5. Placing four $SiPh_2{}^tBuMoO_4$ fragments into the model

The crystal structure of this example has inversion symmetry. However, the $SiPh_2{}^tBuMoO_4$ fragment lacks an inversion center. It does not have a mirror reflection symmetry, either. A model of this fragment cannot reach its inverted version via pure rotations. So, both the model and its inverted version need to be used. The same set of rotation angles that carry the model into a correct orientation also carry the inverted version into a correct orientation—just imagine both the model and its inverted brother riding the same cartesian system which is rotated into the final orientation. With such considerations, the best guess is that orientations 0 and 1 should be used because there are four fragments.

The first fragment uses orientation 0, with its local origin (which is on the Mo atom) sits at (0.3, 0.3, 0.3) in the unit cell. The second fragment uses orientation 1. The placement of its local origin in the unit cell is optimized by minimizing its pR1 function. The $3^{rd}$ and $4^{th}$ fragments should use the inverted version of the fragment and the orientations 0 and 1, respectively. Their placement in the cell is also optimized. The result is a partial model with all four $SiPh_2{}^tBuMoO_4$ fragments as shown in step 1 in figure 2.

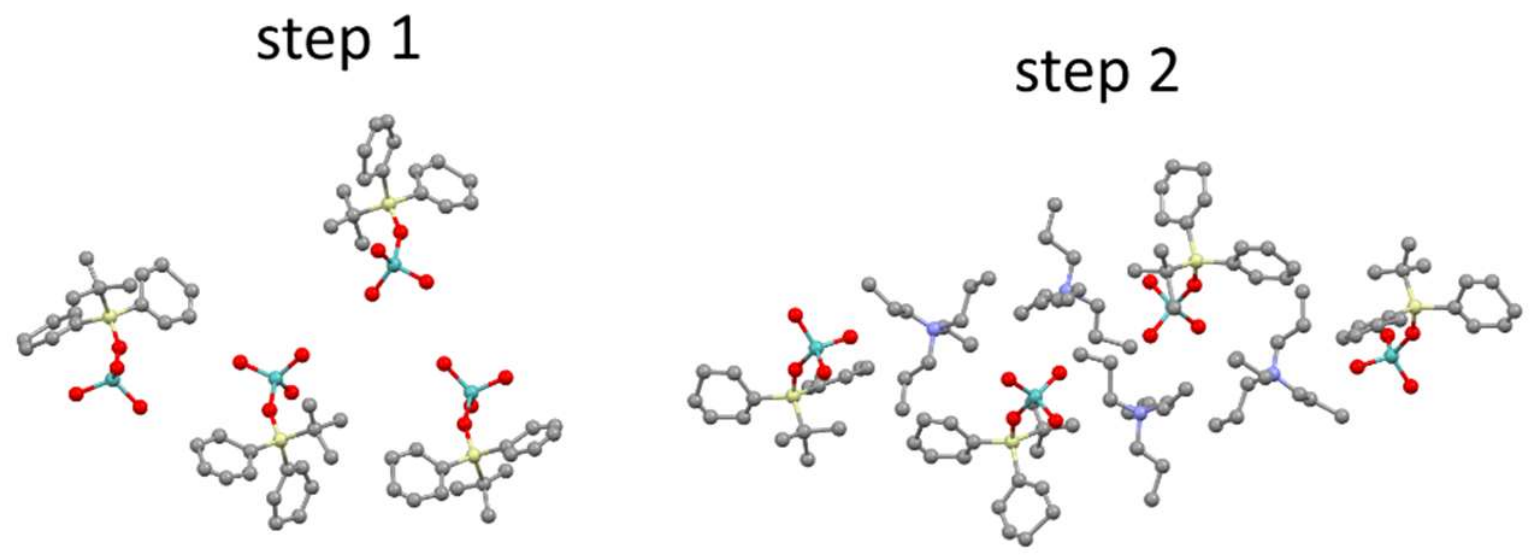

**Figure 2** Steps of assembling the crystal structure of the showcase example by the pR1 method

### 6. Tweaking the partial model to minimize the approximate R1

Note that here the term the approximate R1 (aR1) is used. The aR1 is related to a partial model. There are other atoms that are not included in the partial model. The aR1 is defined via modifying the traditional R1 by removing its dependence on the location of these other atoms. If the partial model contains a single undetermined atom, the aR1 is called the sR1. If the partial model contains an undetermined fragment, the aR1 is called the pR1. Otherwise, all atoms in the partial model have been determined, the aR1 can only be called the aR1.

As the partially finished model was built with idealized components, it does not optimally agree with the experimental data. To optimize it, the position of each atom needs to be tweaked to minimize the aR1. The technical details of how such tweaking is carried out are given in the supporting information. The partial model showing in step 1 of figure 2 has in fact already been tweaked. It is seen that no fragments fall apart after tweak, indicating that the model of the $SiPh_2{}^tBuMoO_4$ fragment was built correctly, and the fragments have been correctly oriented and positioned. One important reason to perform tweaking at this point is that in the next section the residual hkl data will be calculated to determine the orientations of the $N^nPr_4$ fragments. As such residual reflection intensities are weak, the partial model must be optimized to improve the accuracy of these calculated residual reflection intensities.

### 7. Determine the correct orientations of the $N^nPr_4$ fragments

Let one try to use a free-standing $N^nPr_4$ fragment to detect its possible orientations 0, 1, 2, etc. Then use any of these determined orientations to optimally position an $N^nPr_4$ fragment. A positioned fragment usually will bond to some $SiPh_2{}^tBuMoO_4$ fragments. Occasionally the resulting $N^nPr_4$ fragment does not invade other fragments, but it falls apart when the partial model goes through tweaking. Such failure is understandable. The crystal structure of this example contains heavy atoms (the Mo and Si atoms) while the $N^nPr_4$ fragment only contains light atoms (N and C). Trying to use a free-standing $N^nPr_4$ fragment to model the whole structure is doomed to have the fragment trying to match the diffraction power of the heavy atoms. The signals from the target fragments are too weak to be seen.

To overcome this difficulty, residual diffraction intensities should be calculated by deducting away the contributions of the four $SiPh_2{}^tBuMoO_4$ fragments. Technical details of how such residual diffraction intensities are calculated are given in the supporting information. Using the residual hkl data and a free-standing $N^nPr_4$ fragment a set of possible orientations are calculated and are labeled as 0, 1, 2, ...

### 8. Place four $N^nPr_4$ fragments to the model

Even though an $N^nPr_4$ fragment does not have an inversion center, it has mirror reflection symmetry. Therefore, a model of this fragment can reach its inverted version via pure rotations. So, only the model itself needs to be used (no need to use its inverted version), and the best guess is that orientations 0, 1, 2 and 3 should be used. Using each of these orientations an $N^nPr_4$ fragment can be optimally placed in the unit cell by minimizing its pR1 function. In this way, four $N^nPr_4$ fragments are added, one after the other. After adding these fragments, the whole model goes through tweaking to make it optimally agree with the experimental data. Step 2 of figure 2 shows the resulting model of the whole crystal structure after such tweaking. It is seen that no fragment falls apart, which means that the resulting model agrees with the experimental data well.

### 9. Concluding comments

Besides the example shown above, the pR1 method has also been tested over other datasets. Details of two other examples are given in the supporting information. Based on the observations in these examples, the points given in the introduction section have been confirmed. First, in favorable situations the pR1 method can be used to assemble a crystal structure from a few fragments. If the structure contains large fragments like the $SiPh_2{}^tBuMoO_4$ fragment, the correct orientations of these fragments can be detected by using a free-standing form of the fragment and by searching for the pR1 local minima in the orientation space. If a fragment contains heavy atom(s), even if it is small, like the $MoO_4$ fragment, its correct orientation can still be determined in the same way. With orientations known, fragments can be optimally placed in the unit cell by minimizing their pR1 functions. Second, when building a structural model, those fragments that are attached to the growing partial model are easy to add because only their orientations need to be optimized. Third, however, for a structure containing heavy atoms, it is not possible to directly search for orientations of a light-atom-only fragment. In this situation, the correct strategy is to determine the partial model that contains all the heavy atoms first. Then, after tweaking, calculate the residual reflection intensities. Using the residual intensities and a free-standing form of the fragment the orientations of a light-atom-only fragment can then be calculated.

### 10. Computer and software used

A Microsoft Surface Pro 9 [$12^{th}$ Gen Intel® Core™ i7-1255U (2.60 GHz), 32.0 GB installed RAM] was used. Computing time for determining the orientations of one fragment ranged from a few minutes to about one hour. A translational search for one fragment took about one to two hours. Software for the sR1 and pR1 calculations has been coded in Python. The Python codes are available on GitHub (https://github.com/xzhang222/sR1-method). A text file README.md explains how to use the Python codes to do the sR1 and the pR1 calculations.

**Acknowledgements** Professor Igor V. Rubtsov, Professor James P. Donahue and Professor Mark Sulkes are thanked for continued support on the sR1 and pR1 work.

# Supporting information

### S1. Set up an idealized model of the $N^nPr_4$ fragment

This model is shown below together with the choice of the local cartesian system. All bonds have 1.5 Å bond length. All bond angles are 109.5º. All atoms are either in x-y plane or in x-z plane.

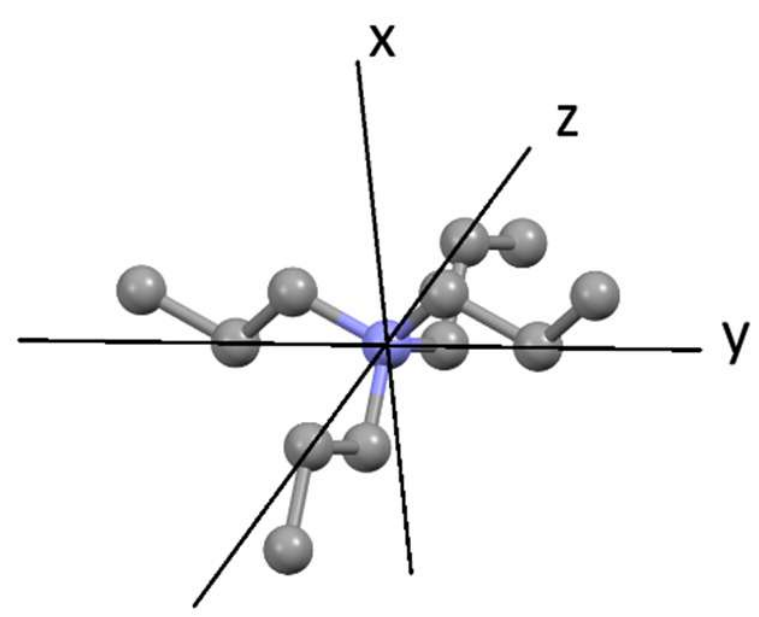

### S2. Tweaking the positions of the atoms in a (partial) model to minimize the aR1

Select a step size 0.4 Å. Randomly select an order of the atoms. For each atom set up a local grid of 27 points, with the original location of the atom at the center of the grid, with one step up and one step down along each cell edge direction. Calculate aR1 values on all 27 grid points. Use the grid point with the lowest aR1 value as the improved location of the atom. Try this on all atoms in the order that has been randomly selected. Finish all these calculations with step size 0.4 Å, then repeat them with step size 0.2 Å, and finally repeat again with step size 0.1 Å. All above counts as one cycle of tweaking. Such cycles are repeated until there is no lowering of the aR1 value.

### S3. Calculating residual diffraction intensities

$F_o$ is the amplitude of the structure factor as derived from the observed reflection intensity (after proper scaling). $F_c$ is the amplitude of the structure factor as calculated from the (partial) model. Let $\delta$ be the residual reflection intensity. Three ways of calculating $\delta$ have been tested:

Choice 1: $\delta = F_o^2 - F_c^2$ if $F_o^2 > F_c^2$; 0 otherwise

Choice 2: $\delta = F_o - F_c$ if $F_o > F_c$; 0 otherwise

Choice 3: $\delta = (F_o - F_c)^2$ if $F_o > F_c$; 0 otherwise

For calculating the orientations of a light-atom-only fragment all these choices work equally well. There are only minor differences in the result, like rotation angles being different slightly, or the same orientation being expressed by totally different rotation angles, etc. So, there is no clear winner. By pure mathematical thinking, choice 1 is selected (as it looks more like calculating residual “intensity”, because $F_o^2$ is proportional to the observed intensity).

## S4. Example 2 of using pR1 to assemble a crystal structure

Dataset code : RAP119. Space group P-1.

(Note that the example for the showcase is example 1; so, this example is called example 2.)

The structure of this example consists of 26 benzene rings. It is possible to search for all the 13 orientations of these benzene rings and assemble this structure by placing all benzene rings one by one. That approach was tried and was successful, but not very efficient. Here a more efficient approach will be used by attacking larger fragments first.

The structure contains four benzene stars. Following is an idealized model of a benzene star with C-C bond length 1.39 Å:

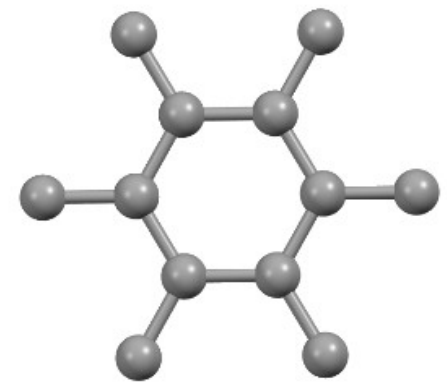

Using a free-standing form the possible orientations of the benzene star are detected and are labeled as 0, 1, 2…

The crystal structure has inversion symmetry, and a benzene star has its own inversion center. So, four benzene stars share two orientations 0 and 1 because the inverted benzene star shares the same orientation with the original benzene star. The first benzene star uses orientation 0 and can be placed with its center at (0.3, 0.3, 0.3) in the unit cell. The 2$^{nd}$ uses orientation 1 and the location of its center within the unit cell being optimized by minimizing its pR1 function. Similarly, the 3$^{rd}$ and 4$^{th}$ use orientations 0 and 1, respectively, and their placement in the cell is also optimized. The result is a partial model with four benzene stars:

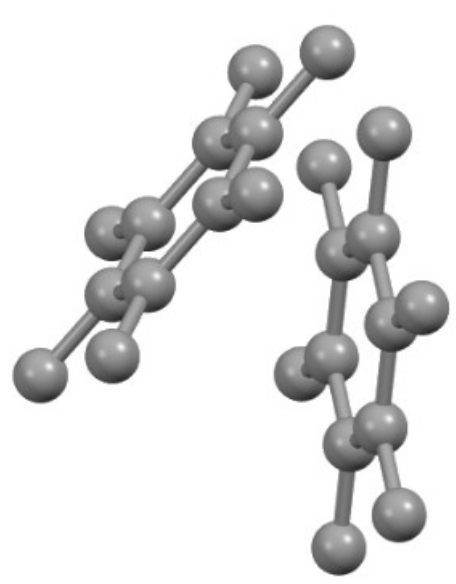

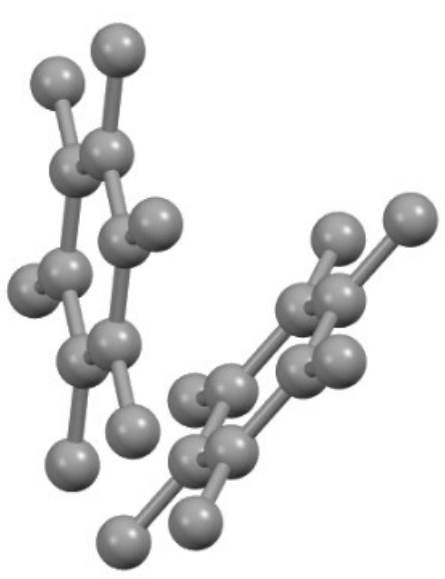

After tweaking they are connected in two pairs:

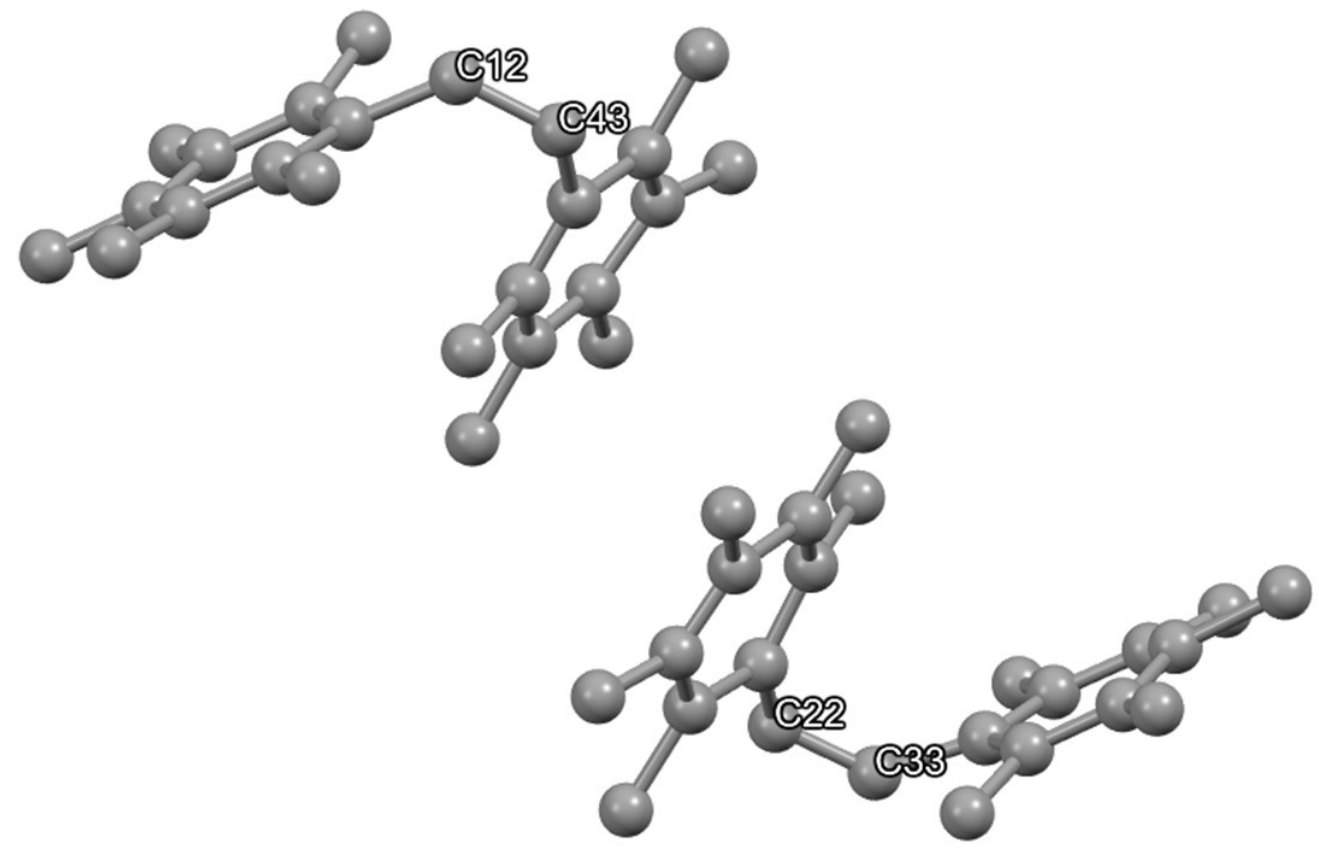

The carbon pair C12-C43 should be completed into a benzene ring; same as C22-C33. Set up an idealized benzene ring model, using C-C bond length 1.39 Å. Set its local origin at middle of one side and delete the two C atoms on this side. The result is a model of $C_4$ fragment. Attach one $C_4$ fragment to the model with its local origin sitting at the center of C12-C43 and its orientation being optimized. Add another one on C22-C33. The result is the following partial model:

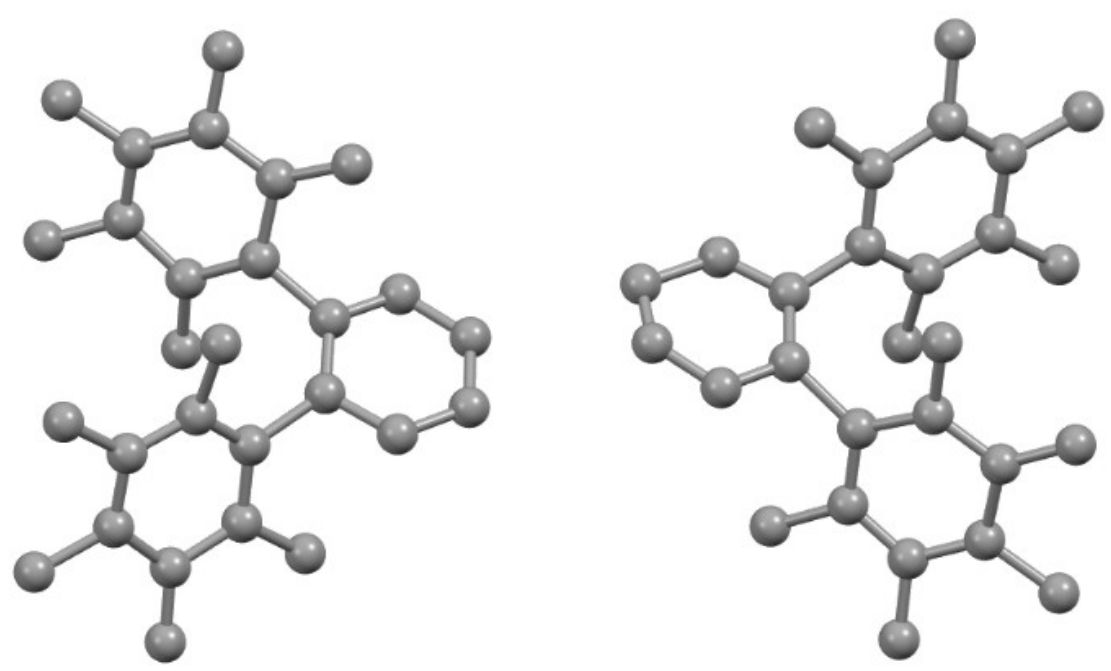

Each single C atom that is sticking out in this partial model should be completed into a benzene ring. Set up an idealized model of a benzene ring with C-C bond length 1.39 Å. Put local origin at one C atom and then delete this atom. The result is a model of a $C_5$ fragment. Add one $C_5$ fragment to the model with its origin sitting on one sticking-out C atom and its orientation being optimized. Do this for all sticking-out C atoms. The result is a completed model. The following is this completed model after tweaking all C atom positions to minimize the aR1 value:

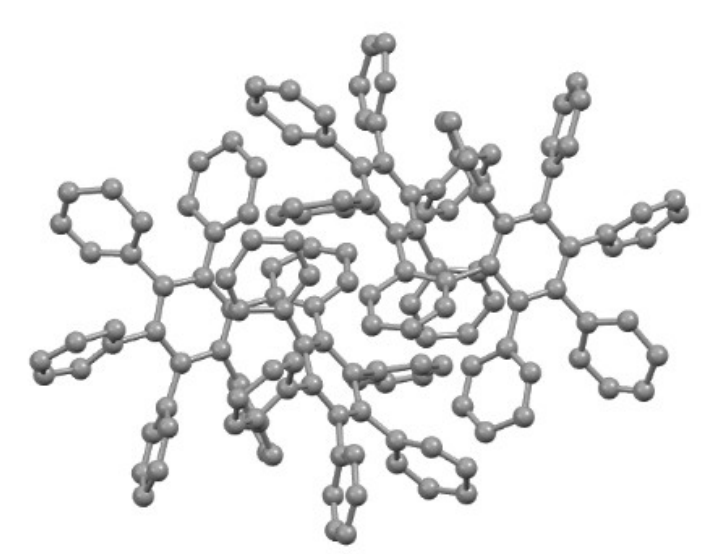

## S5. Example 3 of using pR1 to assemble a crystal structure

Dataset code: JPD1249. Space group P21.

The purpose of this example is to see how heavy atoms interfere with the detection of the orientations of light-atom-only fragments.

There are two S2O2C12 molecules in the structure. The following is a correctly determined structure of one S2O2C12 molecule:

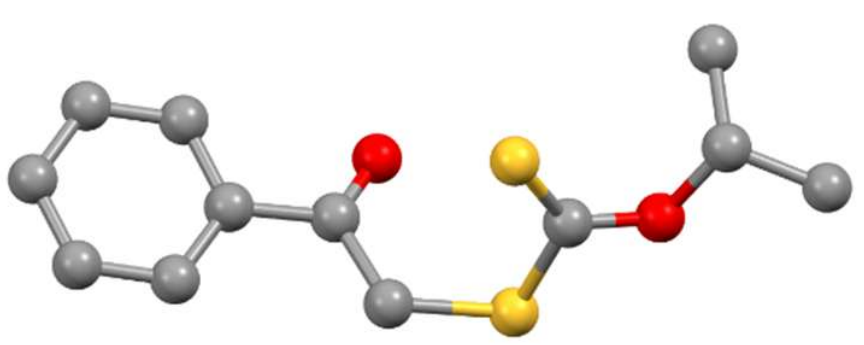

For doing a pR1 calculation experiment, this molecule is cut into three fragments, S2, OC8 and OC4:

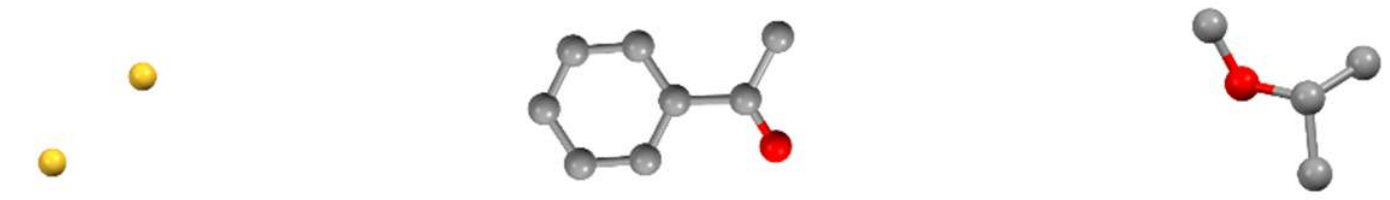

Doing pR1 calculation two S2 can be placed into the cell to get the following partial model:

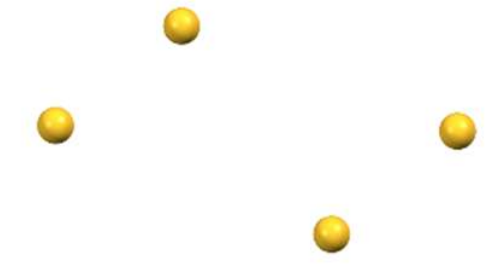

Use a free-standing OC8 to find a set of possible orientations 0, 1, 2, … of the OC8 fragment. Then use orientations 0 and 1 to place two OC8 fragments into the model. However, this does not work, very likely due to the interference of the heavy atoms (i.e. the S atoms):

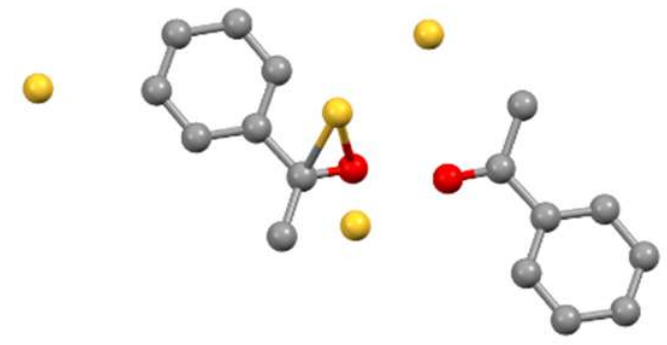

Go back to the partial model with only 4 S atoms and calculate a set of residual reflection data. Using this residual reflection data and a free-standing OC8 to find a new set of possible orientations 0, 1, 2… Try orientations 0 and 1 to place two OC8 fragments, but it is still not right:

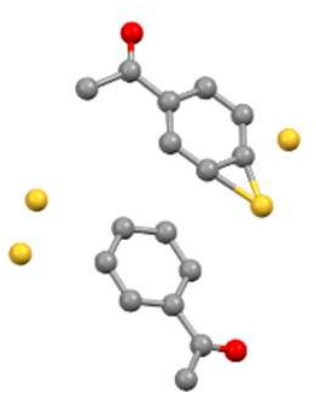

This indicates that the interference persists. Fortunately, upon trying orientations 2 and 3 the result is correct:

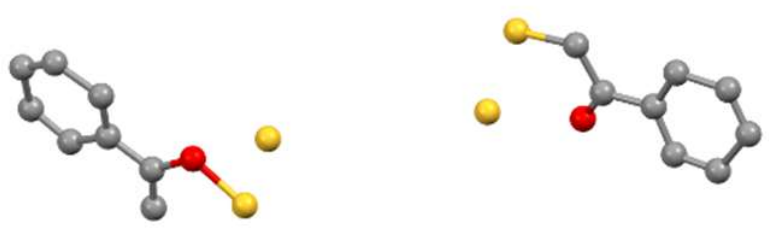

Before trying OC4 fragments, first tweak this partial model and generate a new set of residual reflection intensities. Use the new set of residual intensities and a free-standing OC4 to find a set of possible orientations 0, 1, 2… for OC4. Using orientation 0 one OC4 can be successfully added to the model. However, adding the second OC4 fragment is troublesome. The best is using orientation 3, which, however, is still not acceptable:

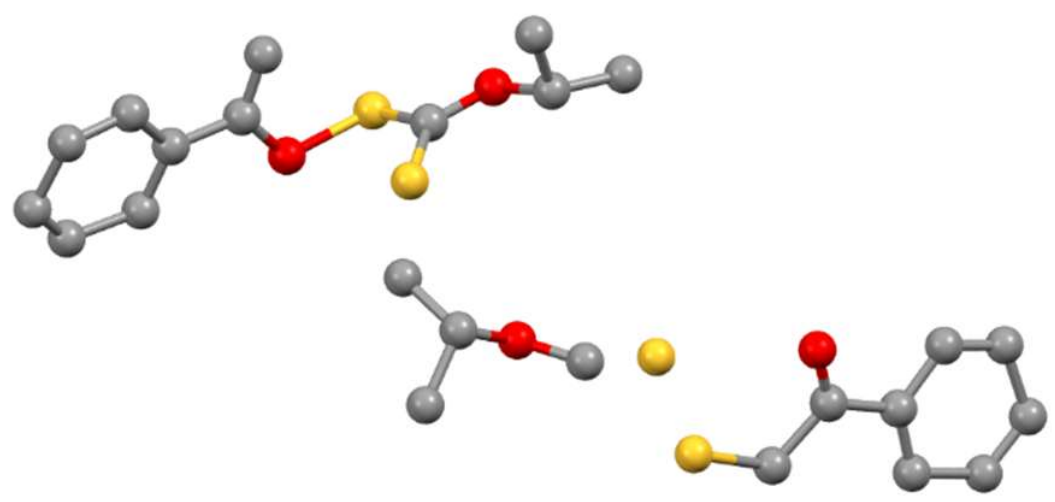

Delete this second OC4 for now and calculate a new set of residual reflections. Use this new set and a free-standing OC4 to determine a new set of possible orientations 0, 1, 2… This time, when trying new orientation 1 the result is finally correct:

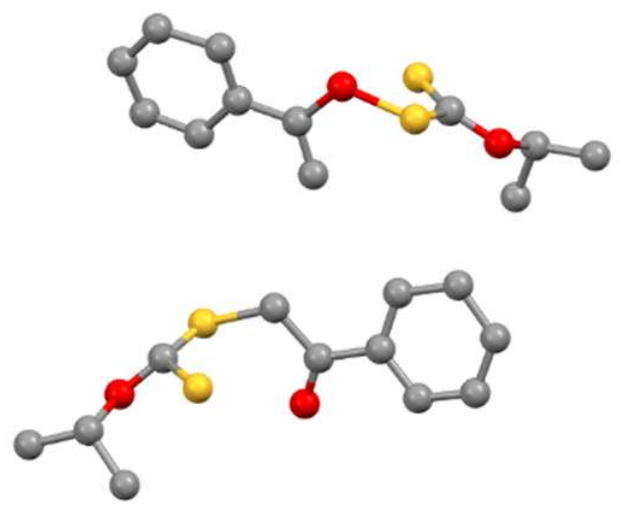

And the following is the final model after tweaking all atom positions:

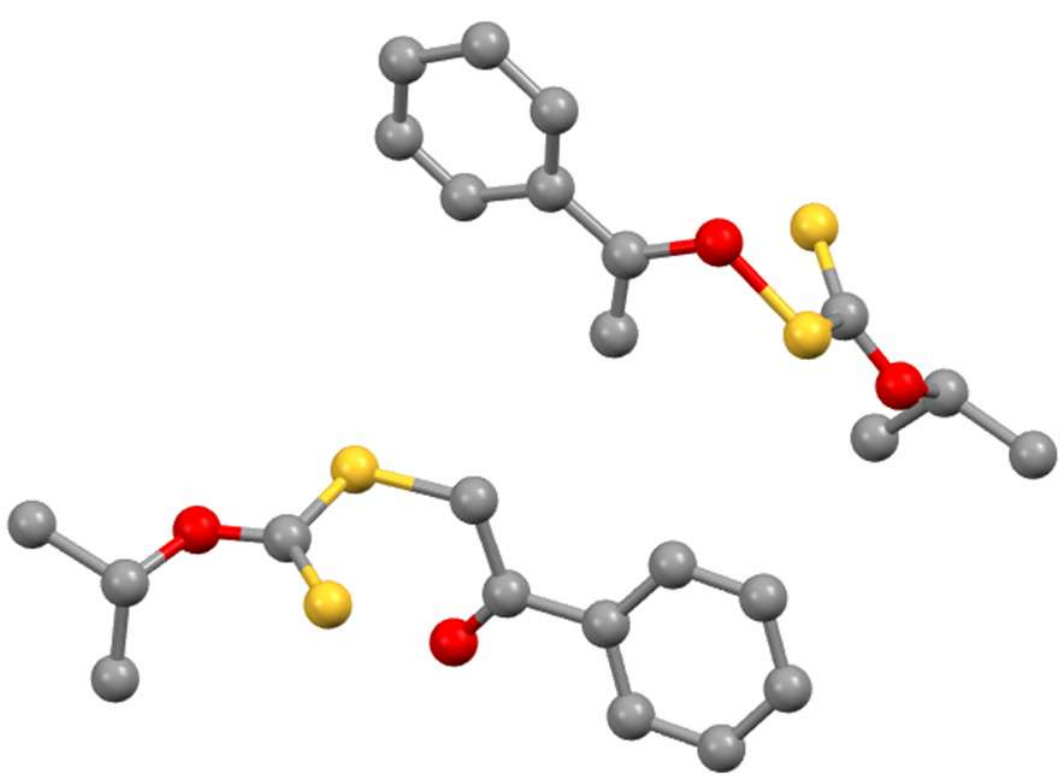

This example demonstrates that the existence of heavy atoms can severely interfere with detecting correct orientations of a light-atom-only fragment. To overcome this difficulty, it is necessary to use residual diffraction intensities.

If super computing power is available such that direct 6-d search can be performed to simultaneously optimize the orientation and position of a fragment, all these troubles are expected to disappear. Of course, correct strategy should still be followed: deal with fragments containing heavy atoms and/or large fragments first. Using 6-d search the pR1 can be a general tool for solving small-molecule crystal structures. For now, it can only solve some favorable structures.